\documentclass[journal]{IEEEtran}
\usepackage{graphicx}
\usepackage{siunitx}

\begin{document}
\title{ Commissioning of the New Multi-layer Integration Prototype of the\\
CALICE Tile Hadron Calorimeter}
%
%

\author{Aliakbar Ebrahimi,~\IEEEmembership{on behalf of the CALICE collaboration}
\thanks{A. Ebrahimi is with the  Deutsches Elektronen-Synchrotron (DESY), Hamburg, Germany (e-mail: aliakbar.ebrahimi@desy.de).}%
}

\maketitle
\pagestyle{empty}
\thispagestyle{empty}

\begin{abstract}
The basic prototype of a tile hadron calorimeter (HCAL) for the International Linear Collider (ILC) has been realised and extensively tested. A major aspect of the proposed concept is the improvement of the jet energy resolution by measuring details of the shower development and combining them with the data of the tracking system (particle flow). The prototype utilises scintillating tiles that are read out by novel Silicon Photomultipliers (SiPMs) and takes into account all design aspects that are demanded by the intended operation at the ILC. Currently, a new 12 layer prototype with about 3400 detector channels is under development. Alternative architectures for the scintillating tiles with and without wavelength-shifting fibres and tiles with individual wrapping with reflector foil are tested as well as different types of SiPMs. The new prototype was used for the first time at the CERN Proton Synchrotron test facility in fall 2014. Additionally, detector modules for the CALICE scintillator-based Electromagnetic Calorimeter (Sc-ECAL), that follow the proposed HCAL electronics architecture, were part of this new prototype. A new multi-layer Data Acquisition System (DAQ) was used for the detector configuration and operation.
\end{abstract}


\section{Introduction}
%
%
%
%
\IEEEPARstart{N}{OVEL} high performance calorimeters for experiments at high energy colliders are studied within the CALICE collaboration \cite{calice}. These detectors are optimised for the Particle Flow approach to calorimetry. Particle Flow Algorithms (PFA) require highly granular calorimeters to reduce confusion arising from particles depositing their energy in close vicinity of each other \cite{thomsonPFA}.

Since on the one hand, to have efficient tracking the PFA demands for a large inner radius of the detector and on the other hand the diameter of the magnet should be as small as possible for cost reasons, a trade-off study is required. The detector design resulting from these studies offers rather limited space in between absorber layers for active material and frontend electronics.

The CALICE Analogue Hadron Calorimeter (AHCAL) is a highly granular sampling calorimeter designed to fulfil the aforementioned requirements. The AHCAL uses \num{3 x 3} \si{\square\centi\meter} plastic scintillating tiles as active material which are read out by novel SiPMs \cite{ahcal1}, \cite{ahcal2}. Analogue/digital front-end ASICs, SPIROC2b, are used to read out the SiPMs \cite{spiroc}. Each ASIC supports up to 36 detector channels and has an analog memory array with a depth of 16 per channel where charge measurement and timing information are stored. A 12-bit Wilkinson ADC is embedded in the ASIC which digitises content of the analog memory. It also allows channel-wise gain adaption to compensate for spreads in channels gains. Together with the equalisation of the light yields by the on-chip bias DACs, this allows for operation of all channels of an ASIC with a single threshold.

Because of the very limited space available, active cooling inside the absorber structure is not possible and power dissipation of the integrated electronics has to be limited as much as possible. This is mainly achieved by the so-called power-pulsing scheme \cite{power_pulsing} in which all active electronic parts of the inner detector are switched off in between the bunch trains of the ILC \cite{ilc} bunch structure.

\begin{figure}[!t]
\centering
\includegraphics[width=3.5in]{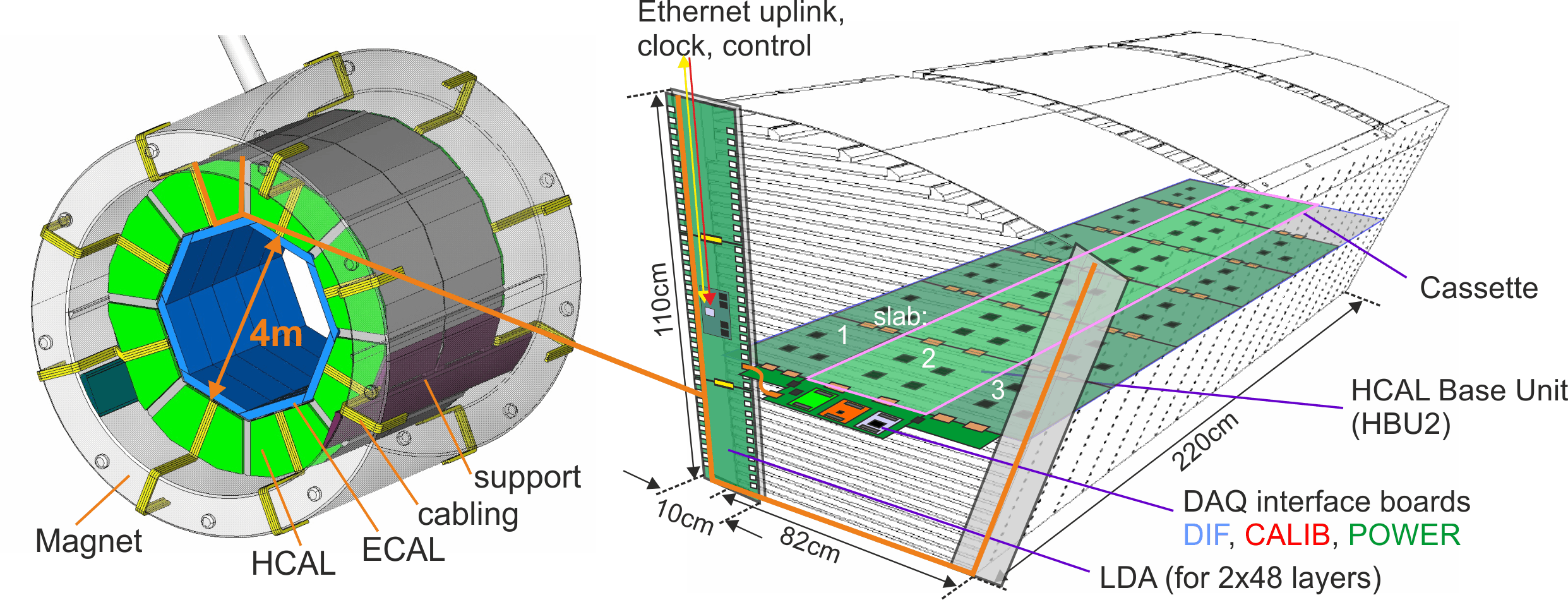}
\caption{Left: A possible barrel structure for calorimeters at the ILC inside the magnet. The green section is the HCAL and the blue section is the electromagnetic calorimeter.
Right:  {1/16} of one half of the AHCAL barrel. One layer of the AHCAL electronics inside the absorber layers is shown.}
\label{fig_barrel}
\end{figure}

\section{Analogue Hadron Calorimeter Concept}

The AHCAL barrel will be divided into 16 sectors around, and 2 subdivision along the beamline as shown in figure \ref{fig_barrel} left. Figure \ref{fig_barrel} right shows the resulting half-sector of the AHCAL. All of the detector interface boards are placed at the two end-faces of the barrel, making them easily accessible for maintenance and service lines \cite{prototype}.

Each detector channel consists of a \num{3 x 3 x 0.3} \si{\cubic\centi\meter}  plastic scintillator tile coupled to a SiPM. Different options for tile design and various types of SiPMs are tested which is described in section IV.

The front-end electronics of each AHCAL layer consist of smaller detector modules called HCAL Base Unit (HBU). Each HBU is \num{36 x 36} \si{\square\centi\meter}  and hosts 144 detector channels. There are 4 SPIROC2b front-end ASICs on each HBU to read out SiPMs. Currently, the third version of this module, HBU3, is available which has improvements over the previous versions. HBU3 has a new circuit for LED drivers of the calibration system which provides improved chanel-to-chanel uniformity. Furthermore, it has improved grounding and reduced voltage drop across one slab (6 HBUs connected in a row).

Each full layer of the AHCAL has a DAQ Interface (DIF) module which provides communication between the fronted electronics and DAQ system. Each layer has a power regulator module and a calibration system control module as well. These three modules are hosted on a Central Interface Board outside of the absorber stack as depicted on figure \ref{fig_barrel} (right).

\section{Data Acquisition System}

\begin{figure}[!t]
\centering
\includegraphics[width=3.5in]{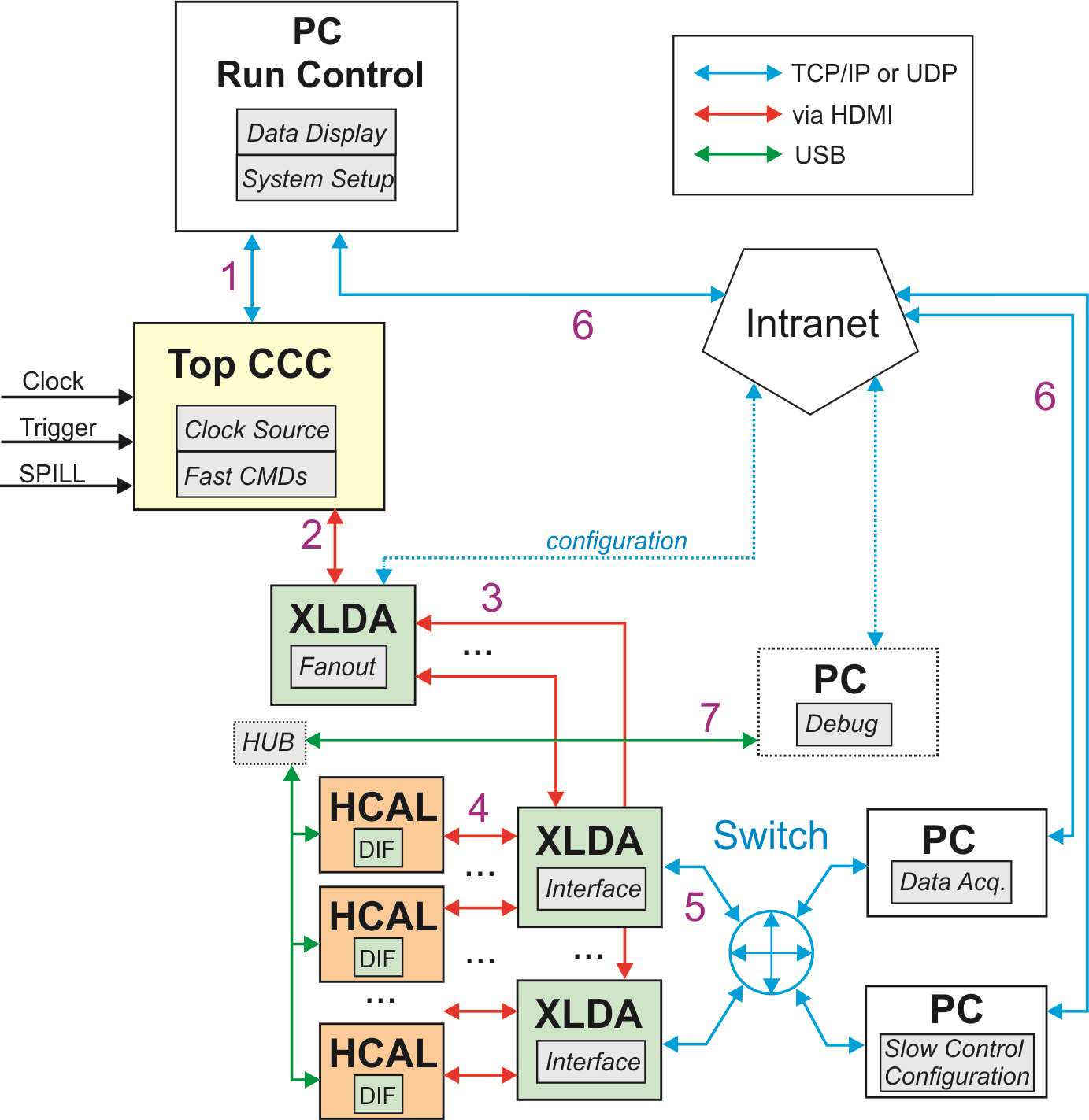}
\caption{Block digram of the AHCAL multi-layer prototype DAQ system}
\label{fig_daq}
\end{figure}

A system overview of the data acquisition system for the AHCAL multi-layer prototype is illustrated in the block diagram of figure \ref{fig_daq}. The new multi-layer DAQ system is based on the original CALICE DAQ concept and is scalable to a full detector.

The Run Control computer is the master of operation and supervises system initialisation, detector configuration and data taking operation. The Run Control software which runs on this PC is realised in a LabView environment which allows for quick and relatively easy modification based on needs during beam test campaigns or system development. Currently the same PC is used for slow control configuration and to store readout data but it is foreseen to have these tasks distributed over extra PCs as shown in the block diagram of figure \ref{fig_daq}. 

The Clock and Control Card (CCC) provides a global clock to the system and guarantees synchronous operation of all detector layers. The new CCC is based on the Xilinx Zynq-7000 All Programmable SoC \cite{zynq} and is realised using a custom mezzanine card attached to a commercially available ZedBoard.

The Link and Data Aggregator (x-LDA) provides interfaces to detector layers and processes and routes data packets being exchanged between various sub-systems. Two different design are available for the x-LDA: a generic 10 port device which can be used in the lab and in smaller setups (mini-LDA), and a larger version which is specifically designed for the AHCAL barrel geometry and supports 96 layers of the detector (figure \ref{fig_barrel}). Both of the x-LDA types are based on the same Xilinx Zynq-7000 All Programmable SoC. Similar to the CCC, the Mini-LDA is is realised using a custom mezzanine card attached to a commercially available ZedBoard. The Wing-LDA has the Zynq-7000 SoC on a commercially available Mars Module as the master chip and uses 4 extra slave FPGAs in order to be able to satisfy the system requirements and serve 96 detector layers. Positioning of the wing-LDA on the AHCAL barrel is shown on figure \ref{fig_barrel}. It is possible to have multiple x-LDA in the current system.

The PCs are communicating with the CCC and x-LDAs over Ethernet links. Standard TCP/IP protocol is used for communication. HDMI cable and connectors are used as physical link between the CCC and the x-LDAs as well as between the x-LDA and the DIFs on detector layers. HDMI is used only as physical layer and no HDMI protocol is implemented. Each HDMI connection provides 5 pairs of wires and each of these pairs are used for differential signalling. Three of the pairs are used for level signals while the remaining two pairs are used for digital serial data transfer. A USB connection is available for laboratory tests and debugging purposes. 

The new DAQ system successfully commissioned and used for the second beam test period at the CERN Proton Synchrotron in fall 2014.

\begin{figure}[!tb]
\centering
\includegraphics[width=3.5in]{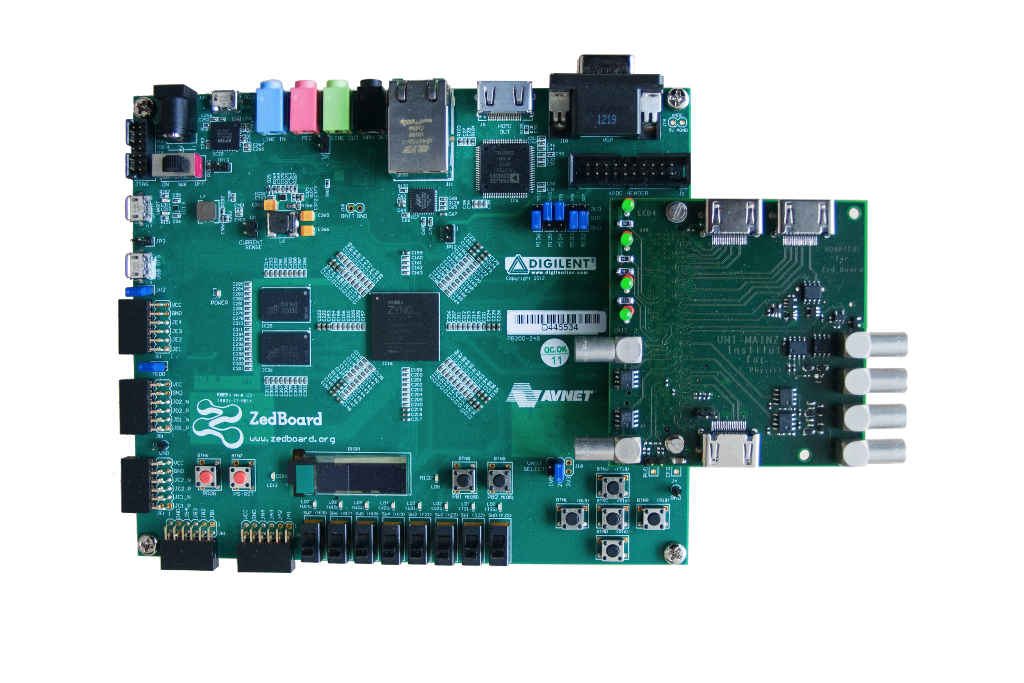}
\caption{A custom CCC mezzanine card attached to a ZedBoard}
\label{fig_ccc}
\end{figure}

\section{Multi-layer Prototype Configuration}

\begin{figure}[!t]
\centering
\includegraphics[width=3.5in]{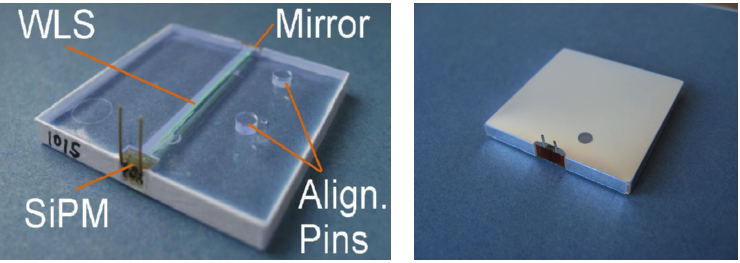}
\caption{Two of the different scintillator tiles used in the AHCAL multi-layer prototype. The tile on the left is an older version with a wavelength-shifting fibre and the tile on the right is a new design by university of Hamburg which is wrapped in reflective foil and has no wavelength-shifting fibre.}
\label{fig_tile}
\end{figure}

Currently the multi-layer prototype consists of 12 modules. These modules differ from one another in HBU version, tile design or SiPM type. Most recent modules are using blue-sensitive SiPMs and therefore have no wavelength-shifting fibres which leads to a simpler process for tile production. Some of the modules are equipped with tiles wrapped in reflector foils (figure \ref{fig_tile}). Various types of SiPMs from different manufacturers are used which have improved characteristics compared to older versions. Number of pixels on these SiPM types varies between 600 to 12000 pixels, which makes it necessary to have specific ASIC settings for each type. Prior to inserting into the stack, various characteristics of each module were studied during the commissioning phase in a laboratory setup and appropriate input DACs and preamplifier values were set. 

The current multi-layer prototype is setup in the EUDET steel absorber structure which has the specifications as planned for the International Large Detector (ILD) barrel at the ILC. For two test beam periods in fall 2014 at the CERN PS, the AHCAL module were combined with 3 CALICE Scintillator Electromagnetic Calorimeter (Sc-ECAL) base units (EBU) \cite{ecal} which have the same electronic architecture as HBUs. In the stack, the 3 EBUs were in the front followed by 8 single HBU layers and 4 layers of 2x2 HBUs at the end. 


One CCC and two Mini-LDAs were used in the setup. One of the Mini-LDAs had 8 layers connected and the remaining 7 layers were connected to the second Mini-LDA. The Mini-LDAs were connected to 2 identical HDMI ports on the same CCC. A gigabit Ethernet switch was used to create a local area network and link the Run Control PC to the CCC and the Mini-LDAs.

During the second beam period, one layer of the CALICE  Silicon Electromagnetic Calorimeter (Si-ECAL) was setup in front of the absorber stack. Since Si-ECAL has a different electronics architecture and data acquisition system, the aim was to run both systems using a combined DAQ system. Synchronous operation of both systems was achieved by using the CCC of the scintillator DAQ as the global clock provider and master of operation and the EUDAQ software framework to start and stop data taking and data storage of the two systems.

\begin{figure}[!t]
\centering
\includegraphics[width=3.5in]{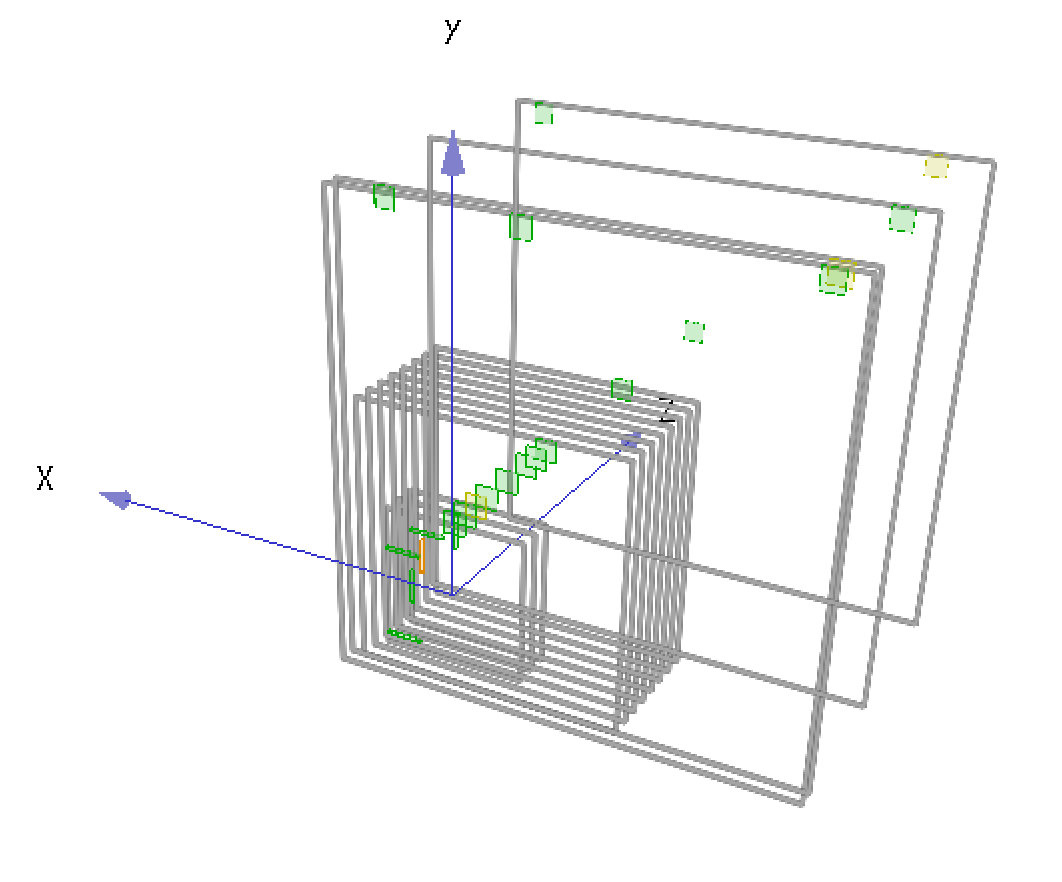}
\caption{Configuration of the layers in the EUDET absorber stack at the CERN PS test beam. A reconstructed muon track in the prototype is shown.}
\label{fig_muon}
\end{figure}

Figure \ref{fig_muon} shows a reconstructed muon track in the multi-layer prototype.

\section{OUTLOOK}

After successful commissioning of the new detector modules and the new multi-layer DAQ, the prototype will be operated in test beams at DESY and at the CERN Super Proton Synchrotron for further technological tests such as power-pulsing as well as physics studies with different absorber material.

New active layers will be added to the prototype after undergoing standard commissioning procedure. 

\section{Conclusions}
A new multi-layer integration prototype of the Analogue Hadron Calorimeter, AHCAL, based on scintillator tiles and SiPMs is under development within the CALICE collaboration. A new data acquisition system is developed and successfully commissioned which is scalable to a full detector. The current version of the prototype is validated in a combined setup with the CALICE scintillator Electromagnetic Calorimeter  at the CERN Proton Synchrotron, addressing critical electrical and mechanical design issues that are relevant for operation of the calorimeter at the ILC.

\vfill


%

\end{document}